\documentstyle[prd,aps,preprint]{revtex}
\tightenlines
\input epsf.tex
\def\DESepsf(#1 width #2){\epsfxsize=#2 \epsfbox{#1}}

\begin{document}

\draft
\preprint{\vbox{
\hbox{UMD-PP-00-006}
\hbox{OSU-HEP-99-08}
\hbox{FERMILAB-PUB-99/214T} }}
\title{Neutrino Masses and Oscillations in Models with Large Extra
Dimensions}

\author{ R. N. Mohapatra$^1$\footnote{e-mail:rmohapat@physics.umd.edu},
S. Nandi$^{2,3}$\footnote{Summer visitor at Fermilab;
e-mail:shaown@okstate.edu} and 
A. P\'erez-Lorenzana$^{1,4}$\footnote{e-mail:aplorenz@Glue.umd.edu} }

\address{$^1$ Department of
Physics, University of Maryland, College Park, MD, 20742, USA\\
$^2$ Department of Physics, Oklahoma State University,
Stillwater, OK, 74078, USA\\
$^3$ Fermi National Accelerator Laboratory, Batavia, IL. 60510, USA\\
$^4$  Departamento de F\'\i sica, 
Centro de Investigaci\'on y de Estudios Avanzados del I.P.N.\\
Apdo. Post. 14-740, 07000, M\'exico, D.F., M\'exico.}
\date{July, 1999}
\maketitle
\begin{abstract}
{We discuss the profile of neutrino masses and mixings in models with
large
extra dimensions when right handed neutrinos are present in the branes
along with the usual standard model particles. In these models, string
scale must be bigger than $10^{8}$ GeV to have desired properties for the
neutrinos at low energies. The lightest neutrino mass is zero and there is
oscillations to sterile neutrinos that are different from other models
with the bulk neutrino.}
\end{abstract} 

\vskip0.5in

\section{Introduction}
Last year has seen an explosion of interest 
and activity in theories with large extra dimensions 
\cite{many1,many2,lykken,many3,ddg,many}. It has been realized that extra
dimensions almost as large as a millimeter could apparently be
hidden from many extremely precise measurements that exist in
particle physics. What makes such an idea exciting is the hope that the
concept of hidden space dimensions can be probed by collider
as well as other experiments in not too distant future. Furthermore, it
brings the string scale closer to a TeV in some scenarios, making details
of string physics within reach of experiments.
On the theoretical side, recent developments in strongly
coupled string theories have given a certain amount of credibility to such
speculations in that the size of extra dimensions can be
proportional to the string coupling and therefore in the context
of strongly coupled strings such large dimensions are quite plausible.

This new class of theories have a sharp distinction from the conventional
grand unified theories as well as old weakly coupled string models in that
in the latter case, most scales other than the weak scale and the QCD
scale were assumed to be in the range of $10^{13}$ to $10^{16}$ GeV. That
made it easier to understand observations such as small neutrino
masses and a highly stable proton etc. Now that most scales are allowed to
be small in the new models, the two particularly urgent questions that
need to be answered are why is the proton stable and why are neutrinos so
light. It has been speculated that proton stability may be understood by
conjecturing the
existence of U(1) symmetries that forbid the process. On the other hand no
such simple argument for understanding small neutrino masses seems to
exist. The familiar seesaw mechanism\cite{grsyms} is not implementable in
simple versions of these models where all scales are assumed to be low. So
understanding lightness of neutrinos is therefore a challenge to such
models.
 
One approach to this problem discussed in recent
literature\cite{dienes,dvali} is to use neutrinos that live in the bulk
(they are therefore necessarily singlet or sterile with respect to normal
weak interactions) and to observe that
their coupling to the known neutrinos in the brane is inversely
proportional to the square-root of the bulk volume. The neutrino mass
(which is now a Dirac mass) can be shown to be $\sim \frac{h v
M^*}{M_{P\ell}}$, where $M^*$ is the string scale. If the string scale is
in the TeV range, this leads to a neutrino mass of order $10^{-4}$ eV or
so. This therefore explains why the neutrino masses are small. A key
requirement for small neutrino masses is therefore that the string scale
be in the few TeV range\footnote{By choosing the Yukawa coupling to
be smaller, the string scale could of course be pushed higher}.
Furthermore, these models also generally
predict oscillations between the modes of the bulk neutrino and the known
neutrinos. Thus if one wants an explanation of the
solar neutrino deficit via neutrino oscillations, this implies that there
must be at least one extra
dimension with size in the micrometer range\cite{dvali,pospel}. The latter
provides an interesting connection between neutrino physics, gravity
experiments searching for deviations from Newton's law at sub-millimeter
distances as well as possible collider search for TeV string
excitations. It then follows that larger values of the string
scale would jeopardize this simple explanation of
the small neutrino mass. So the extent that larger values of string scale
are also equally plausible as the TeV value, one might search for
alternative ways to understand the small neutrino masses.

 It is the goal
of this paper to outline such a scenario and study its consequences. The
particular example we consider illustrates this scenario
in models with large extra dimensions and generic brane-bulk
picture for particles, where the string scale is necessarily bigger ($\geq
10^{8}$ GeV or so) and solar or atmospheric neutrino oscillations require
at least one extra dimension be in the micrometer range. The new
ingredient of the class of models we discuss here is that we include
the right-handed neutrino in the brane and consider the gauge interactions
to be described by a left-right symmetric model. In these models,
the left-handed neutrinos are not allowed to form mass terms with the bulk
neutrino due to extra gauge symmetries; instead it is only the
right-handed neutrino which is allowed to form mass terms with the bulk
neutrinos. This leads to a different profile for the neutrino 
masses and mixings.
In particular, we find that in this model, the left-handed neutrino is
excatly massless whereas the bulk sterile neutrinos have masses related to
the size of the extra dimensions and it will be of order $\sim
10^{-3}$ eV if there is at least one large extra
dimension with size in the micrometer range. A key distinguishing feature
of this model from the existing ones 
is that the string scale is now necessarily much larger than a TeV. We
also find that 
the pattern of the neutrino oscillations is different from the previous
case.

As mentioned, the minimal gauge model where our scheme is realized is the
left-right
symmetric models where the right handed symmetry is broken by the doublet
Higgs bosons $\chi_R (1,2,1)$. The notation we follow is that the
three numbers inside the parenthesis correspond to the quantum numbers
under $SU(2)_L\times
SU(2)_R\times U(1)_{B-L}$. We do not need supersymmetry for our discussion
and will therefore work within the context of nonsupersymmetric left-right
models.

To set the stage for our discussion,
let us start with a review of the neutrino mass mechanism in models with
large extra dimensions discussed in Ref.\cite{dienes}. The basic idea is
to include the
coupling of bulk neutrino $\nu_B( x^{\mu}, y)$ (which is a standard model
singlet) to the standard model
lepton doublet $L(x^{\mu}, y=0)$. The Lagrangian that is responsible for
the neutrino masses in this model is:
\begin{eqnarray}
{\cal S} = \kappa \int d^4x \bar{L} H \nu_B(x, y=0) + \int d^4x dy
\bar{\nu}_B(x,y)\Gamma^5\partial_5 \nu_{B}(x,y) + h.c.
\end{eqnarray}
Writing the four component spinor $\nu_{B}\equiv \left(\begin{array}{c}
\nu^1_B \\ i\sigma_2\nu^{2*}_B \end{array}\right)$, we get for the
neutrino mass matrix:
\begin{eqnarray}
 (\bar{\nu}_{eL}  \bar{\nu}'_{BL})\left(\begin{array}{cc}
\kappa v &\sqrt{2} \kappa v\\ 0 & \partial_5
\end{array}\right)\left(\begin{array}{c}\nu_{0B} \\
\nu'_{BR}\end{array}\right)
\end{eqnarray}
This is a compact way of writing the KK excitations along the fifth
dimension. 
Notice that the $\sqrt{2}$ in  the off-diagonal term appears to compensate
the different normalization of the zero mode in the Fourier expansion of
the bulk field in
terms of $\sin ny/R$ and $\cos ny/R$ ($R$ being the
radius of the fifth dimension). Also, notice that 
only the last terms  may couple
to the pure four dimensional fields, and  
we always may choose these modes to
have  positive KK masses.
In the above equation $\kappa$ embodies the features of the
string scale and the radius of the extra dimension i.e. $\kappa\simeq
\frac{M^*}{M_{P\ell}}$.
For $\kappa v \ll \mu_0 $ where $\mu_0=R^{-1}$, 
we get the mixing of the $\nu_e$ with
the bulk modes to be 
\begin{eqnarray}
tan \theta_{eB} \approx \frac{\sqrt{2}\kappa v \partial_5}{\partial^2_5
- \kappa^2 v^2} 
\end{eqnarray}
Substituting the eigenvalues of the operator $\partial_5= n\mu_0$, we
get for the n-th KK excitation a mixing $tan\theta_n\approx
\frac{\xi}{n}$ where $\xi\approx \frac{\sqrt{2}\kappa v}{\mu_0}$. This 
expression is same as in \cite{dienes,dvali}.

Important point to note here is that since $\nu_e$ has a mass of $\kappa
v$, present neutrino mass limits lead to an upper limit on $\kappa$ and
hence on the string scale. For instance, if we choose the
present tritium decay bounds\cite{mainz} of $m_{\nu_e}\leq 2.5$
eV, we get $M^*\leq 10^{7}$ GeV. This bound gets considerably strengthened 
if we further require that the solar
neutrino puzzle be solved via $\nu_e-\nu_s$ oscillation (where we have
called the typical excited models of the bulk neutrino as ``sterile
neutrinos''). The reason is that this requires $\Delta m^2\simeq
10^{-5}$ eV$^2$ and in the absence of any unnatural fine tuning, we will
have to assume that that $m_{\nu_e}\simeq m_{\nu_s}\simeq 10^{-3}$
eV. This implies that $M^*\leq 10$ TeV and the bulk radius is given by
$R^{-1}\simeq 10^{-3}$ eV implying $R\simeq 0.2 $ mm. Another prediction
of this model is that the mixing of the $\nu_e$ with the bulk neutrinos
goes down like $(1/n)$ where $n$ denotes the level of Kaluza-Klein
excitation.

Let us now proceed to the new case we are considering. The part of the
action relevant to our discussion is given by:
\begin{eqnarray}
{\cal S}= \int d^4x [\kappa\bar{L}\chi_L \nu_B(x, y=0) +
\kappa \bar{R}\chi_R
\nu_B(x, y=0)+h \bar{L}\phi R] + \int d^4x dy
\bar{\nu}_B\Gamma^5\partial_5\nu_B + h.c.
\end{eqnarray}
where $L^T=(\nu_{eL}, e_L)$ and $R^T=(\nu_{eR}, e_R)$ and $\phi$ is the
bidoublet Higgs field that breaks the elctroweak symmetry and gives mass
to the charged fermions. We
assume that the $SU(2)_R$ gauge group is broken by $<\chi^0_R>= v_R$ with
$<\chi^0_L>=0$\footnote{In general, in the left-right model, there is a
coupling of the form $M_0\bar{\chi}_L\phi\chi_R$ which induces a vev for
the $\chi^0_L$ field of order $\frac{M_0 v_{wk}}{M_{str}}$, which for a
choice of $M_0=v_{wk}$ gives a $<\chi^0_L>\sim 0.1$ MeV. We assume $M_0$
to be smaller so that this contribution to the neutrino masses is
negligible. It could also be that the symmetries of the charged fermion
sector are such that they either forbid such a coupling or give such a vev
to $\phi$ that this term does not effect the value of the potential at
the minimum. We thank the referee for pointing this out.} and the
$SU(2)_L\times U(1)_Y$ is broken by $Diag<\phi>=
(v, v')$. The profile of the neutrino mixing matrix in this case is
given by
\begin{eqnarray}
(\bar{\nu}_{eL}~\bar{\nu}_{0BL}~  \bar{\nu}'_{BL})\left(\begin{array}{cc}
hv & 0\\ \kappa v_R & 0 \\\sqrt{2}\kappa v_R &  \partial_5
\end{array}\right)\left(\begin{array}{c}\nu_{eR} \\
\nu'_{BR}\end{array}\right)
\end{eqnarray}
where $hv$ is the usual Dirac mass term present in the models
with the seesaw mechanism and normally assumed to be of order of typical
charged fermion masses (we will also make this plausible assumption
that $hv$ is of the order of a few MeV's for the first generation which
will be the focus of this article).

We can now proceed to find the eigenstates and neutrino mixings. First
point
to note is that this matrix being $2\times 3$ has one zero eigenvalue
corresponding to the state\cite{wyler}
\begin{eqnarray}
\nu_0 = (\cos\zeta ~\nu_{eL} - \sin\zeta ~\nu_{0BL})
\end{eqnarray}
where $sin\zeta =\frac{hv}{\sqrt{(\kappa v_R)^2+(hv)^2}}$. 
If we want the lightest eigenstate to be predominantly the electron
neutrino so that observed universality in charged current weak interaction
is maintained, we must demand that $\kappa v_R\gg hv$. Since $\kappa\simeq
\frac{M^*}{M_{P\ell}}$, this constraint will imply a constraint on the
string scale $M^*$.

Let us see under what circumstances this condition is satisfied. Since the
Dirac mass $hv\simeq$ few MeV's, we would like $\kappa v_R \gg$ few MeV's.
Let us assume that there is one extra dimension with large size
(of order milli-meter and denoted by $R_1$) and all other extra
dimensions have very small sizes, assumed to be equal. Then the observed
strength of gravitational interaction implies the relation
\begin{eqnarray}
M^2_{P\ell}={ M^{*}}^{n+2}R^{n-1}R_1
\label{mpl}
\end{eqnarray}
where $R_1$ is the largest dimension and the rest of the R's are small as
required by neutrino physics. For simplicity let us identify the right
handed symmetry breaking scale, the string scale and the inverse radii of 
the small dimensions $R^{-1}$. Then we have the
approximate relation that
\begin{eqnarray}
\kappa\simeq \frac{1}{\sqrt{M^{*n}R^{n-1}R_1}}\simeq \frac{M^*}{M_{P\ell}}
\end{eqnarray}
$\kappa v_R\gg $ few MeV (say 100 MeV), implies that $M^*\simeq
v_R\approx R^{-1}\geq 10^{8.5}$
GeV. In fact it is not hard to see that to satisfy the relation in 
Eq. (\ref{mpl}),
the radii of the ``small'' compact dimensions must be also of order
${M^*}^{-1}$ and that of the large dimension is of course in the
sub-millimeter range. While this is a generic possibility, one can of
course make many variations on this general theme. The results of this
paper are not effected by such variations. Thus it appears that our scheme
prefers a high string scale in contrast with the earlier
proposal\cite{dienes}.

We can now look at the rest of the neutrino spectrum arising from the 
KK excitations of the bulk mode as well as the right handed  neutrino.
They can be studied by looking at the ``$2\times 2$'' matrix after
extracting the zero mode discussed above. Defining the orthogonal
combination to $\nu_0$ as $\tilde{\nu}_{0L}$, we have:
\begin{eqnarray}
(\bar{\tilde{\nu}}_{0L}~  \bar{\nu}'_{BL})\left(\begin{array}{cc}
\kappa v_R & 0\\\sqrt{2}\kappa v_R &  \partial_5
\end{array}\right)\left(\begin{array}{c}\nu_{eR} \\
\nu'_{BR}\end{array}\right)
\end{eqnarray}
where $\tilde{\nu}_{0L}= \cos\zeta ~\nu_{0BL}+\sin\zeta ~\nu_{eL}$ and we
have used the approximation that $\kappa v_R\gg hv$. In terms of this
short handed notation, the characteristic equation of the Dirac mass
matrix above is  easily computed to be
\begin{equation}
(m^2 - \partial_5^2)\left[m^2 - \kappa^2 v_R^2 + 
{2 m^2 \kappa^2 v_R^2 \over  \partial_5^2 - m^2}\right] = 0 ,
\end{equation}
This, in a manner similar that noted by Dienes et al.\cite{dienes} 
translates into the transcendental equation
\begin{eqnarray}
m_n = \pi \kappa^2 v_R^2 R \ \cot(\pi m_n R)
\end{eqnarray}
where $m_n$ is the mass of the n-th KK state. The 
tower of eigenstates, symbolically denoted by $\nu_{nL}$, are exactly 
given by
\begin{equation}
\nu_{nL} = {1\over \eta_n}\left[ {\tilde{\nu}}_{0L} + 
{\sqrt{2} m_n^2 \over m^2 - \partial_5^2} ~ \nu'_{BL} \right],
\label{nun}
\end{equation}
with the normalization factor, $\eta_n$, given as 
\begin{equation}
\eta_n^2 = 1 + \sum_{k=1}^\infty {2 m_n^4 R^4\over (k^2 - m_n^2R^2)^2} .
\label{etan}
\end{equation}

As long as $\kappa v_R R \gg 1$, all the mass eigenvalues, $m_n$ upto
$m_n\sim \kappa v_R$ satisfy $cot(\pi m_n R)\approx 0$. 
Therefore, the masses for all these states are $\sim(2n-1)/2R$. Thus the
effect of the mixing of the brane righthanded neutrinos is to shift the
bulk neutrino levels below and near $\kappa v_R$ downward. This is similar
to what was noted in the case without the $\nu_R$ in the
brane\cite{dienes} for all states.
On the other hand, using the same equation (Eq. (11)) it is easy to see
 that  the states much heavier than $\kappa v_R$ have masses
$\sim n/R$, which also means that they basically  decouple from  the light
sector. For the eigenstates in the middle, those with masses just 
beyond $\kappa v_R$, their mixing with the lightest states 
strongly suppress their
contribution to $\tilde\nu_{0L}$, by an amount $\sim 1/\kappa v_R R$, 
as it could be checked from Eq. (\ref{etan}), 
by summing over those elements for
which $k\ll \kappa v_R R$.

   There are, of course, similar results for the right handed
states which involve the right handed bulk neutrinos. In fact, 
they are identified
as the negative solutions to the characteristic equation, since the same
mass matrix gives the masses of both left and right handed 
sectors and those are degenerate.

For discussing neutrino oscillation, the left handed eigenstates are the
only ones relevant. For this purpose let us write down the weak
eigenstate $\nu_e$ in
terms of the mass eigenstates through their mixing into 
$\tilde \nu_0$. Since 
$\nu_e= c_\zeta ~\nu_0 + s_\zeta ~\tilde\nu_0$,
with $c_\zeta$ ($s_\zeta$) standing for $\cos\zeta$ ($\sin\zeta$),
the survival probability of $\nu_e$ oscillations reads
\begin{equation}
P(\nu_e\rightarrow\nu_e(t))= c_\zeta^4 + 
s_\zeta^4 ~P(\tilde\nu_0\rightarrow\tilde\nu_0(t)) + 
2 ~c_\zeta^2 ~ s_\zeta^2 ~ Re \langle \tilde\nu_0|\tilde\nu_0(t)\rangle , 
\end{equation}
where $P(\tilde\nu_0\rightarrow\tilde\nu_0(t))$ is the corresponding 
``survival probability'' for $\tilde\nu_0$. In terms of the mass
eigenstates
\begin{equation}
\tilde\nu_0(t) = \sum_{n=1}^N {e^{i (m_n^2 t/2E)}\over \eta_n} \nu_n,
\label{nutilde}
\end{equation}
where we have cut off the sum by $N\approx \kappa v_R R$, explicitly
decoupling the heavy eigenstates. This is justified by the arguments given
above. For the remaining states, one can show that
$\eta_n^2\approx {\pi^2\over 8}(2n-1)^2$ which follows from the following
exact expression for $\eta^2_n$:
\begin{eqnarray}
\eta^2_n~=~1+{1\over 4}\csc^2(2\pi m_n R)\left[2\pi^2 m^2_n R^2 + \pi sin
(2\pi m_n R) + 2 cos (2\pi m_n R) -2\right]
\end{eqnarray}
It means that the main
contribution to $\tilde\nu_0$ comes  from the lightest mode, which is
expected since the bulk zero mode is its main  original component. 
The final
survival probability after the neutrino traverses a distance L in
vacuum can be written down as
\begin{equation}
P_{ee}(E) = 
1 - 4\ c^2_\zeta\ s^2_\zeta \sum_{n=1}^N {1\over  \eta_n^2}
\sin^2\left({m_n^2 L\over 4E}\right) - 
4\ s^4_\zeta\ \sum_{k<n}^N {1\over  \eta_n^2 \eta_k^2}
\sin^2\left[{(m_n^2-m_k^2)L\over 4E}\right].
\end{equation}
Therefore, the oscillation length is given by
$L_{osc}=  E R^2_1\sim (E/MeV)\times 5\cdot 10^4$m for $R_1= 0.1 ~mm$.
This value for the oscillation length is right in the domain of
accessibility of the KAMLAND experiment\cite{kamland}.
In Fig. 1 we  present
the survival probability as a function of distance for specific values of
$\kappa v_R$ and $h v$.
Also, we present the probability for the  
early proposal of Ref.~\cite{dvali} in Fig. 2 for comparision. 
Notice that the 
differences in both profiles come, basically, from two facts: first, 
the masses of the light bulk 
modes have been shifted down by ${1\over 2} \mu_0$ in the present case.
Such an effect is absent in the approach followed in the former
case\cite{dienes}.
This is reflected in a (four times) larger  oscillation length. 
Secondly, the mixing of $\nu_e$ with bulk states is different in our case
than Ref.\cite{dienes,dvali} though for large values of $n$ they coincide.
 
The averaged probability is now obtained in a stgraightforward manner to
be $\overline{P_{ee}} = c_\zeta^4 + {2\over 3} s_\zeta^4$, which is
smaller than the two neutrino case with the same mixing angle, although,
for $\zeta\ll 1$ it approaches the former  result 
$\overline{P_{ee}}\approx 1 - 2\zeta^2$. 
Moreover, we may average over all the modes, but the lowest frequency one,
to get
\begin{equation}
P_{ee}(E) = c_\zeta^4 + {2\over 3} s_\zeta^4 + 
{16\over\pi^2} c_\zeta^2\ s_\zeta^2\ \left[1 - 2 \sin^2\left({\mu_0^2
L\over 16 E}\right)\right].
\end{equation}
Hence, the depth of the  oscillations is now of the order of 
${32\over \pi^2} c_\zeta^2 s_\zeta^2$.

In conclusion, we have presented a new way to understand small neutrino
masses in models with large extra dimensions by including the righthanded
neutrino among the brane particles. In addition to several quantitative
differences from earlier works which have only then standard model
particles in the brane, we find that the string scale in the new class of
models is necessarily larger. However, we need one extra
dimension to be of large size if we want to solve the solar neutrino
problem as well have other meaningful oscillations between the known
neutrinos with the bulk neutrinos.

We have not discussed the matter effect and the implications for solar
neutrino puzzle nor have we discussed the astrophysical constraints on
our scenario. We hope to return to these topics subsequently. But we have
noted that multiple ripple effect induced by the KK modes in the $\nu_e$
survival probability that was noted for the earlier case in
Ref.\cite{dvali} remains in our case too.

 Secondly,
we have focussed only on the first generation neutrinos; but there is no
obstacle in principle to extending the discussion to the other
generations. Clearly, as in the case of Ref.\cite{dvali}, attempting to
solve the atmospheric neutrino puzzle by $\nu_{\mu}$-$\nu_{bulk}$
oscillation would require that we make the bulk radius $R_1$ smaller.

Our overall impression is that while it is possible to understand the
smallness of the neutrino mass using the bulk neutrinos without invoking
the seesaw mechanism and concommitant high scale physics such as $B-L$
symmetry, a unified picture that clearly incorporates all three
generations with preferred mixing and mass patterns\cite{bilenky} is yet
to come whereas in existing ``four-dimensional'' models there exist
perhaps a surplus of ideas that lead to desirable neutrino mass textures.
>From this point of view, the neutrino physics study in models with extra
diensions is still in its infancy and whether it grows into adulthood
depends much on the direction that these extra dimensional models take in
the coming years.

\vskip1em

{\it Acknowledgements.}  
The work of RNM is supported by a grant from the National
Science Foundation under grant number PHY-9802551. The work of SN is
supported by the DOE grant  DE-FG03-98ER41076.
The work of APL is supported in part by CONACyT (M\'exico). RNM would like
to thank the Institute for Nuclear Theory at the University of Washington
 for hospitality and support during the last stages of this work. SN
gratefully acknowledges the warm hospitality and support during his visit
to the University of Maryland particle theory group when this work was
started and to the Fermilab theory group where it was completed.



\begin{figure}
\centerline{
\epsfxsize=250pt
\epsfbox{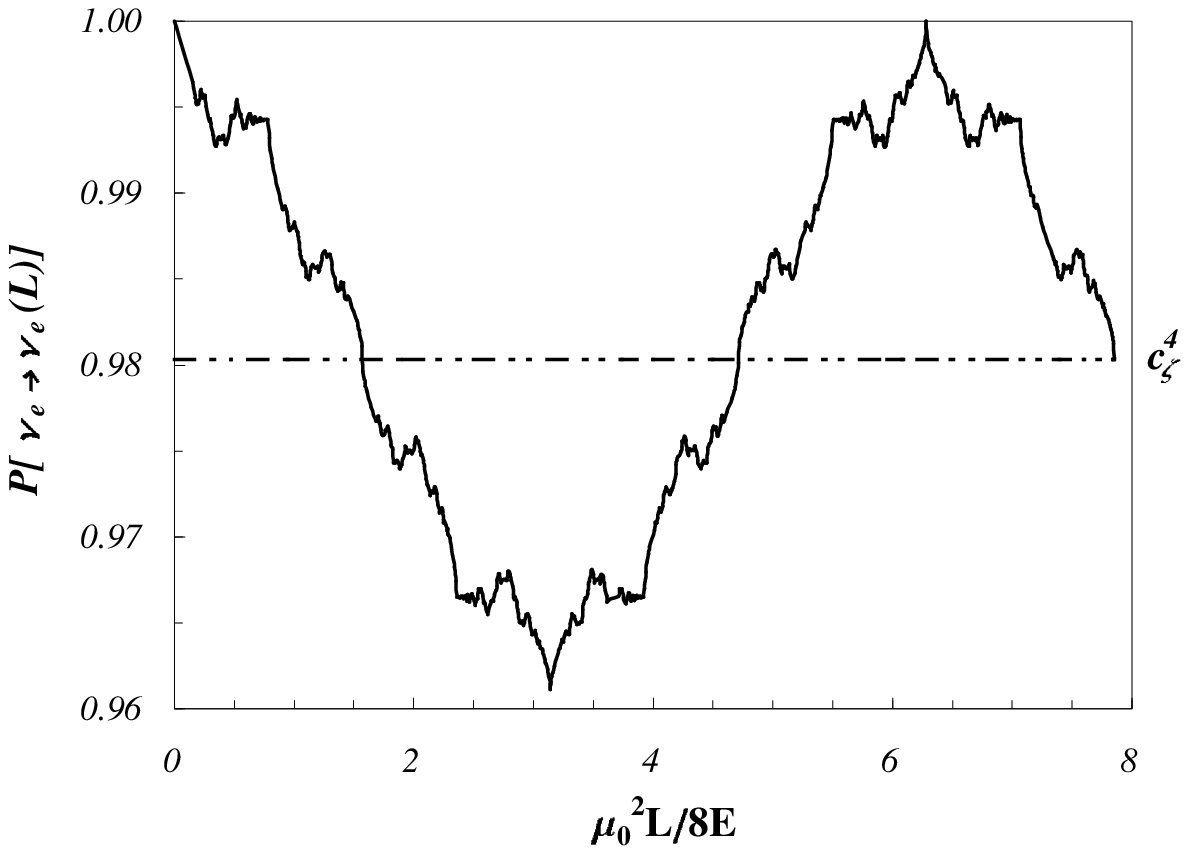}
}
\caption{
Profile of the survival probability for $\nu_e$ oscillations
obtained for $\kappa v_R = 100$ MeV and $hv=10$ MeV. The
horizontal line corresponds to the value of $\cos^4\zeta$ 
}
\end{figure}
\vskip1em

\begin{figure}
\centerline{
\epsfxsize=250pt
\epsfbox{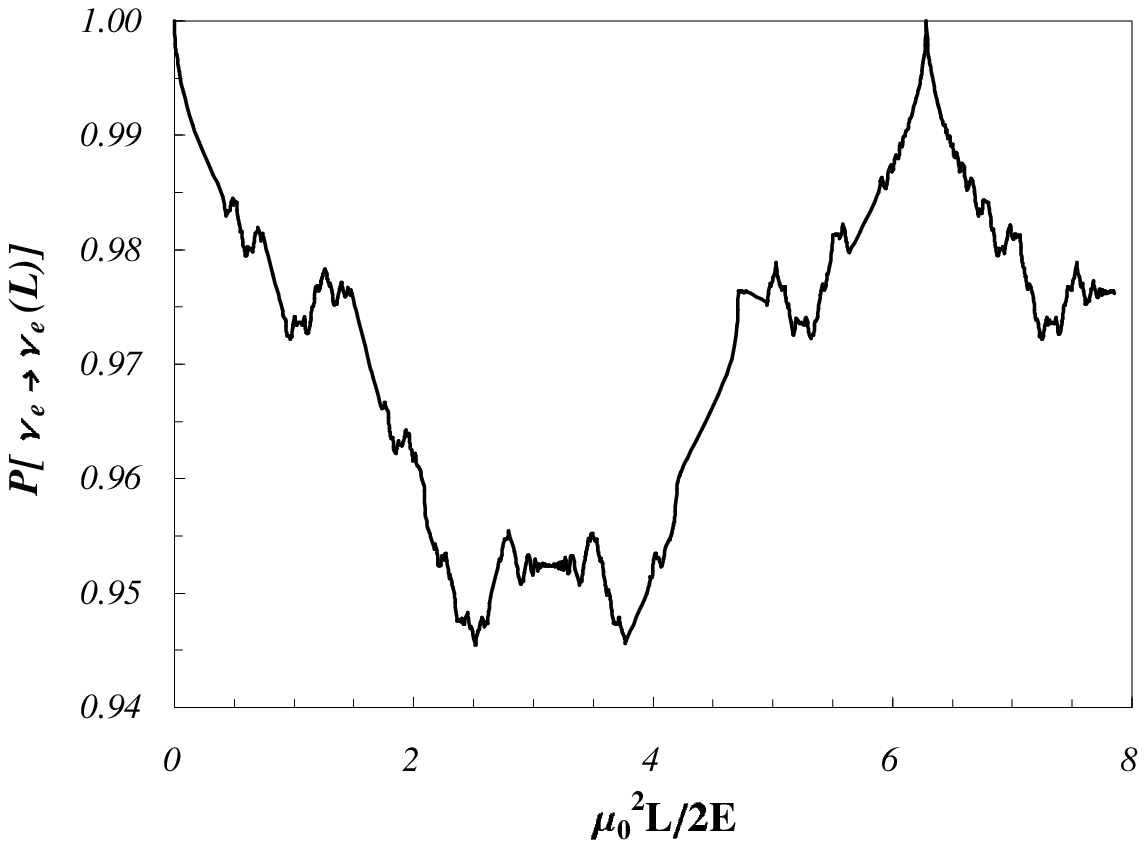}
}
\caption{
Profile of the survival probability for $\nu_e$ oscillations
for the model of first Ref. in [8]
using $\xi = 0.1$. Notice that the argument is
augmented by a factor of four with respect to the case in the previous
figure. 
}
\end{figure}


\begin{thebibliography}{99}

\bibitem{many1}   I. Antoniadis, Phys. Lett. {\bf B246} (1990) 377; I.
Antoniadis, K. Benakli and M. Quir\'os, Phys. Lett. {\bf B331} (1994) 313.

\bibitem{many2} E. Witten, Nucl. Phys. {\bf B 471}, 135 (1996);
 P. Horava and E. Witten, Nucl. Phys. {\bf B475} (1996) 94.

\bibitem{lykken} J. Lykken, \prd {\bf 54} (1996) 3693.

\bibitem{many3} N. Arkani-Hamed, S. Dimopoulos and  G. Dvali,
\pl {\bf B429} (1998) 263; \prd {\bf 59} (1999) 086004;
I. Antoniadis, S. Dimopoulos, G. Dvali,
Nucl. Phys. {\bf B516} (1998) 70.

\bibitem{ddg}  K.R. Dienes, E. Dudas and  T. Gherghetta, 
\pl {\bf B436} (1998) 55;  Nucl. Phys. {\bf B537} (1999) 47.  

\bibitem{many} 
K. Benakli, hep-ph/9809582,\pl {\bf  B447} (1999) 51;
P.Nath and M. Yamaguchi, hep-ph/9903298; hep-ph/9902323;
M. L. Graesser, hep-ph/9902310;
M. Masip and A. Pomarol, hep-ph/9902467;
T. Banks, A. Nelson and M. Dine, JHEP 9906 (1999) 014;
D. Ghilencea and  G.G. Ross, \pl {\bf B442} (1998) 165;
Z. Kakushadze, Nucl. Phys. {\bf B548} (1999) 205;
C.D. Carone, \pl {\bf B454} (1999) 70;
A. Delgado and M. Quir\'os, hep-ph/9903400.
P. H. Frampton and A. R\v{a}sin, hep-ph/9903479;
G. Giudice, R. Rattazzi and J. Wells, Nucl. Phys. {\bf B544} (1999) 3; 
E. Mirabelli, M. Perelstein and M. Peskin, \prl {\bf 82} (1999) 2236;
T. Han, J. Lykken and R. J. Zhang, \prd {\bf 59} (1999) 105006;
J. L. Hewett, \prl {\bf 82} (1999) 47656; 
P. Mathews, S. Raychaudhuri and K. Sridhar, \pl {\bf B450} (1999) 343; 
T. G. Rizzo, \prd {\bf 59} (1999) 115010; 
K. Aghase and N. G. Deshpande, \pl {\bf B456} (1999) 60;
K. Cheung and W. Y. Keung, hep-ph/9903294;
T. Taylor and G. Veneziano, Phys. Lett. {\bf B212} (1988) 147; 
C. Burgess, L. Iba\~nez and F. Quevedo, \pl {\bf B447} (1999) 257;
A. P. Lorenzana and R. N. Mohapatra, hep-ph/9904504; 
K. Huitu and T. Kobayashi, hep-ph/9906431.
D. Dumitru and S. Nandi, hep-ph/9906514;
C. Balazs, H. -J He, W. W. Repko, C. P. Yuan and D. A. Dicus,
hep-ph/9904220; T. Han, D. Rainwater and D. Zepenfield, hep-ph/9905423; W.
J. Marciano, hep-ph/9903451; A. Delgardo and M. Quiros, hep-ph/9903400; H.
C. Cheng, B. Dobrescu and C. Hill, hep-ph/9906327; T. Rizzo and J. Wells,
hep-ph/9906234.

 \bibitem{dienes}   K.R. Dienes, E. Dudas and  T. Gherghetta,
hep-ph/9811428; N. Arkani-Hamed, S. Dimopoulos, G. Dvali and J.
March-Russell, hep-ph/9811448.

\bibitem{dvali} G. Dvali and A.Yu. Smirnov, hep-ph/9904211.

\bibitem{pospel} A. Faraggi and M. Pospelov, hep-ph/9901299; A. Das and
O. C. W. Kong, hep-ph/9907272.

\bibitem{grsyms}  M. Gell-Mann, P. Ramond and R. Slansky, in {\it
Supergravity}, eds. P. van Niewenhuizen and D.Z. Freedman (North
Holland 1979); T. Yanagida, in Proceedings of {\it Workshop on
 Unified Theory and Baryon number in the Universe}, eds.
O. Sawada and A. Sugamoto (KEK 1979);  R. N. Mohapatra and
G. Senjanovi{\'c}, Phys. Rev. Lett. {\bf 44}, 912 (1980).

\bibitem{mainz} C. Weinheimer, Invited talk at the workshop on {\it Lepton
Moments}, Heidelberg, June (1999); V. Lobashev et al. Phys. Lett {\bf B}
(to appear).

\bibitem{wyler} L. Wolfenstein and D. Wyler, Nucl. Phys. {\bf B 218}, 205
(1983).

\bibitem{kamland} F. Suekane et al. preprint TOHOKU-HEP-97-02 (1997).

\bibitem{bilenky} For recent reviews see, S. Bilenky, C. Giunti and W.
Grimus, hep-ph/9812360, to appear in {\it Prog. in Particle and Nucl.
Phys., vol. 43}; B. Kayser, P. Fisher and K. Macfarland, hep-ph/9906244
(to appear in the {\it Ann. Rev. of Nucl. and Part. Sc.}, 1999).

\end{thebibliography}
\end{document}